\documentclass{article}
\usepackage[utf8]{inputenc}
\usepackage[english]{babel}
\usepackage{newtxtext,newtxmath}
\usepackage[dvipdfmx]{graphicx}
\usepackage{authblk}
\usepackage{url}
\usepackage{CJKutf8}
\begin{document}
\title{You are what you eat: A social media study of food identity}

\author[$\dagger$,$\ddagger$]{Kazutoshi Sasahara\thanks{Correspondence should be addressed to K.S. (sasahara@nagoya-u.jp)}}
\affil[$\dagger$]{Department of Complex Systems Science, Nagoya University, Furo-cho, Chikusa-ku, Nagoya 458-8601, Japan}
\affil[$\ddagger$]{JST, PRESTO, Kawaguchi, Japan}
\date{}

\maketitle

\begin{abstract} 
Food preferences not only originate from a person's dietary habits, but also reflect personal values and consumer awareness. 
This study addresses ``food identity'' or the relationship between food preferences and personal attributes based on the concept of ``food left-wing'' (e.g., vegetarians) and ``food right-wing'' (e.g., fast-food lovers) by analyzing social data using information entropy and networks. 
The results show that food identity extends beyond the domain of food: The food left-wing has a strong interest in socio-environmental issues, while the food right-wing has a higher interest in large-scale shopping malls and politically conservative issues. 
Furthermore, the social interactions of food left-wing and right-wing factions show segregated structures, indicating different information consumption patterns. 
These findings suggest that food identity may be applicable as a proxy for personal attributes and offer insights into potential buying patterns.
\end{abstract}

\section{Introduction}
\section*{Introduction}
In today's age of gluttony, we are overwhelmed with information about food.
Food is available at a moment's notice in supermarkets, and we continually see advertisements for food online. 
With almost unlimited options, the choice of what to eat and what not to eat depends less on
biological aspects, such as individual survival or likes and dislikes, but rather it reflects the values of the individual~\cite{Guptill2017}. 
Brillat-Savarin, a French gastronome famously stated, ``Tell me what you eat and I will tell you what you are''~\cite{Brillat-Savarin2009}. 
This indicates that food was believed to be linked with identity. 
Although several studies have examined this aspect (e.g.,~\cite{Laufer2015}), much remains understudied.

There are terms typifying the relationship between food and political ideologies that appear in sociological discourse in American culture, such as ``Starbucks people'' and ``Coors beer people''~\cite{Watanabe2008}. 
These terms are meant as ironic signifiers of the metropolitan intelligentsia (liberal and Democratic Party-supporting people), who buy expensive coffee and read the New York Times at Starbucks versus rural people (conservative and Republican Party-supporting people), who drink inexpensive canned Coors beer while watching live broadcasts of American football. 
Other terms that similarly parody these two camps' lifestyles and political ideologies are ``latte liberals'' and ``bird-hunting conservatives''~\cite{DellaPosta2015}.

In Japan, it remains to be seen whether such stark stereotypes would apply, but there is prior work describing the political and dietary sensibilities of Japanese people using the terms ``food left-wing'' and ``food right-wing''~\cite{Hayamizu2013}. 
Proponents of ``food left-wing'' are those who pursue natural food and are health-conscious, typically vegetarians and vegans. 
In contrast, the ``food right-wing'' group generally consumes available food products and enjoys eating fast food.
Hayamizu (2013) describes how food preferences (food left-wing and right-wing) speak to political attitudes in Japan. 

If food preferences reflect the values of people, then these preferences are also likely closely tied not only to political ideology but also to other personal attributes. 
If that were the case, food preferences could be a ``mirror'' reflecting latent consumption preferences and attitudes.

We study ``food identity'' based on the concept of the food left-wing and right-wing in order to determine whether this concept is useful in gaining insights into personal attributes. 
To this end, we analyze social data from Twitter, in which a large amount of food-related information is spontaneously posted and shared. 
Twitter is a better data source for our purse, because its use is better motivated in Japan given the high level of penetration in the country~\cite{hale_global_2014}.
Although there are many previous researches about food-related social data analyses~\cite{Fried2014,Abbar2015,Mejova2015}, little attention has been paid to food identity as a proxy for personal values and consumer awareness. 
This is the main focus of this study.

\section*{Data and Methods}

\subsection*{Data Collection}
The official Twitter Search application programming interface (API)~\cite{TwitterAPI} was used to create a crawler to harvest social data from the site and collected two datasets (Fig. 1)~\footnote{Alternative choice could be the Twitter Streaming API, but it has several issues: e.g., It often does not work properly for non-space separated languages (e.g., Japanese) and it is limited to 1\% of the full Twitter Firehose. Thus, we decided to use the Twitter Search API.}.
We ran the crawler three times a day and thus obtained almost all of the tweets that contained keywords of interest described below. 

\begin{figure}[t]
\centering
 \includegraphics[clip,width=0.9\textwidth]{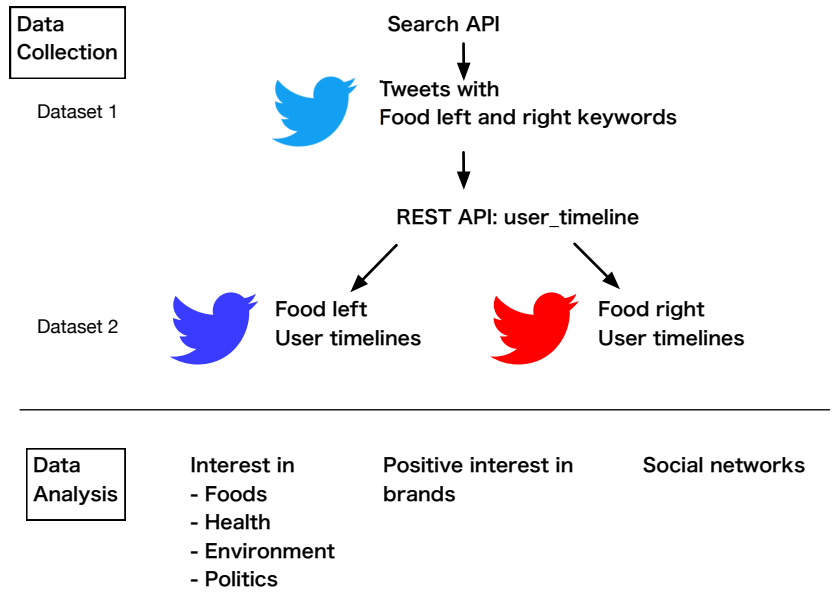}
 \caption{Schematic illustration of data collection and analysis.}
 \label{fig1}
\end{figure}

We used 18 food-related keywords from Shimizu (2013) to identify the food left-wing and right-wing (see Table 1). 
Tweets containing these keywords were collected for a period of approximately one month beginning in December 2016. 
This process yielded 650,900 Japanese tweets containing keywords referencing food left-wing tendencies, and 3,141,527 Japanese tweets containing keywords referencing food right-wing tendencies (Dataset 1). 
Dataset 1 was then used to identify users making 30 or more total tweets containing any food left-wing keywords (using any of the keywords once or more per day, on average). 
The same process was performed for those making tweets containing food right-wing keywords. We excluded users making 30 or more total posts containing both food left-wing and right-wing keywords because they were deemed to not have a specific food preference. 
This process yielded 1,233 users making food left-wing tweets and 5,010 users making food right-wing tweets. 
This research treated the former group as food left-wing and the latter as food right-wing.

\begin{table}[h]
\caption {Keywords used in collecting Dataset 1}
 \begin{tabular}{|l|l|}
 \hline
 Food left  & \begin{tabular}[c]{@{}l@{}}\begin{CJK}{UTF8}{ipxm}地産地消\end{CJK} (local production and consumption),\begin{CJK}{UTF8}{ipxm}スローフード\end{CJK} (slow food), \\\begin{CJK}{UTF8}{ipxm}道の駅\end{CJK} (roadside markets), \begin{CJK}{UTF8}{ipxm}ベジタリアン\end{CJK} (vegetarian), \\\begin{CJK}{UTF8}{ipxm}有機農法\end{CJK} (organic farming), \begin{CJK}{UTF8}{ipxm}オーガニック\end{CJK} (organic),\begin{CJK}{UTF8}{ipxm}マクロビ\end{CJK} (macrobiotic), \\\begin{CJK}{UTF8}{ipxm}ベジフェス\end{CJK} (vegetable festivals),\begin{CJK}{UTF8}{ipxm}ファーマーズマーケット\end{CJK} (farmers' markets)\end{tabular} \\ \hline
 Food right & \begin{tabular}[c]{@{}l@{}}\begin{CJK}{UTF8}{ipxm}ファストフード / ファーストフード\end{CJK} (fast food),\\\begin{CJK}{UTF8}{ipxm}ジャンクフード\end{CJK} (junk food), \begin{CJK}{UTF8}{ipxm}冷凍食品\end{CJK} (frozen foods), \\\begin{CJK}{UTF8}{ipxm}メガ盛り\end{CJK} (super-sized),
 \begin{CJK}{UTF8}{ipxm}デカ盛り\end{CJK} (massive portions),
 \\\begin{CJK}{UTF8}{ipxm}栄養ドリンク\end{CJK} (energy drinks), \begin{CJK}{UTF8}{ipxm}コンビニ\end{CJK} (convenience stores),
 \\\begin{CJK}{UTF8}{ipxm}揚げ物\end{CJK} (fried foods)\end{tabular}                                          \\ \hline
 \end{tabular}
\end{table}

Timelines for all users identified with Dataset 1 were obtained: 3,655,936 tweets from the food left-wing and 15,091,255 tweets from the food right-wing (Dataset 2). 

\subsection*{Data Analysis}
We computed the frequencies of keywords listed in Table 1 and later using texts from Dataset 2. 
To compare the food left-wing and right-wing groups, we constructed 10 bootstrapped samples in each group with 1000 resamplings, used for the measurements described below. 

\subsubsection*{Measuring collective interests}
The frequency of a keyword is often used as an indicator of the aggregate level of interest in a topic on social media, but this poses one major problem. It does not consider how many individuals actually
used the keyword. 
For example, in an extreme case, the most frequently used keyword could be used by a single user. 
To avoid this, we measure keyword entropy (\(H_{k}\)) below as an indicator of collective interest in a keyword by using normalized entropy~\cite{Tria2015}:
\begin{equation}
H_{k} = - \sum_{x \in U}^{}P_{k}(x)\log_{2}P_{k}(x)/\log_{2}N. \notag
\end{equation}
Here, $U$ refers to a set of users, with $P_{k}(x)$ being the probability of a post with the keyword $k$ made by user $x$, and $N$ is the total number of users, with $0 \leq H_{k} \leq 1$. 
As is clear in the definition, when many users make posts including keyword $k$, then $H_{k}$ increases.

To compare keyword entropy between the food left-wing and right-wing groups, we employ the following formula, which is often called the laterality index ($LI$):
\begin{equation}
LI = \frac{H_{k}^{R} - H_{k}^{L}}{H_{k}^{R} + H_{k}^{L}} \notag
\end{equation}
where $R$ represents the food right-wing, and $L$ represents the food left-wing with \(- 1 \leq LI \leq 1\). 
For keyword $k$, a larger $LI$ implies a greater degree of collective interest among food right-wing users; a smaller $LI$ implies a greater degree of collective interest among food left-wing users.

\subsubsection*{Visualizing consumer awareness}
We use association networks~\cite{Sasahara2016} to examine how the food left-wing and right-wing groups are aware of various keywords.
First, the tweet texts from Dataset 2 are segmented into words using MeCab~\cite{Mecab} and the mecab-ipadic-NEologd (a Japanese dictionary)~\cite{neologd}, and then cleaned by removing stopwords (defined in SlothLib~\cite{slothlib}), symbols (e.g., ! and @), and URLs.
Next, the cleaned texts are used as the input to a word embedding method called word2vec~\cite{Mikolov2013a,Mikolov2013}. 
Word2vec can construct lower dimensional vectors that reflect word meanings based on word usage. 
Using the trained word2vec model, we can convert words used within a corpus into vectors, by which semantically similar words would become similar vectors.
We used the Gensim library~\cite{gensim} for the word2vec modeling with the default parameter setting. 

The resulting word vectors are visualized using association networks in the following manner. 
Given a seed word, we list words whose cosine similarity to the seed word vector is greater than the similarity threshold (0.4); we then selected the top 20 most similar words. 
Using the selected words as new seeds, we listed words in a same manner. 
Words obtained by such ``association chains'' are used as nodes. 
If term $w_2$ is selected when term $w_1$ is a seed, $w_1$ and $w_2$ are linked. 
If multiple words are selected when $w_1$ is a seed, then $w_1$ is connected to all these words. 
In this way, we visualize consumer awareness as word associations in tweets. 

\subsubsection*{Social interactions in retweets}
Twitter has the functionality of reposting or retweeting a friend's posts to own followers. 
This leaves a record of how information spreads among users. 
Given that user B retweets a post from user A, A and B can be treated as a node, with the directed link A\(\rightarrow\)B.  
All of the retweets in Dataset 2 can be turned into directed links in this way, allowing for the reconstruction of social interactions in retweets among food left-wing and right-wing users. 
We refer to this as a retweet network, described as \(G = (V,E)\)~\cite{Sasahara2013}. 
Here, \(V\) is a set of the users listed using the above method, and \(E\) is a set of links describing retweet transmissions. 
\(V\) includes users other than the left-wing and right-wing seed users in Dataset 2, but those never retweeted were not included. 
Analyzing the retweet network (\(G\)) allows us to examine structural patterns of information transmissions within and between food left-wing and right-wing groups.

\section*{Results}
\subsection*{Collective interest in food and other keywords}
First, we confirmed whether food left-wing and right-wing users had different food preferences and whether they had other preferences outside of food. 
We computed the frequencies of the keywords in Table 2 for the categories of food, health, socio-environmental issues, and politics that were featured in the Nikkei newspaper from 2015 to 2016 (except for meat, fish, and vegetables)~\footnote{Note that a health freak is a person extremely enthusiastic about health.}. 
The resulting word frequencies were then used to compute keyword entropy to compare the degree of interest in each group. 

\begin{table}[h]
\caption{Keywords related to food, health, socio-environment, and politics}
 \begin{tabular}{|l|l|}
 \hline
 Food        & \begin{tabular}[c]{@{}l@{}}\begin{CJK}{UTF8}{ipxm}肉\end{CJK} (Meat), \begin{CJK}{UTF8}{ipxm}魚\end{CJK} (fish), \begin{CJK}{UTF8}{ipxm}野菜\end{CJK} (vegetables), \\\begin{CJK}{UTF8}{ipxm}遺伝子組み換え\end{CJK} (GM (genetically modified)), \\\begin{CJK}{UTF8}{ipxm}トランス脂肪酸\end{CJK} (trans fatty acid), \\\begin{CJK}{UTF8}{ipxm}カップヌードル\end{CJK} (instant noodles)\end{tabular}  \\ \hline
 Health      & \begin{tabular}[c]{@{}l@{}}\begin{CJK}{UTF8}{ipxm}高カロリー\end{CJK} (High calorie), \begin{CJK}{UTF8}{ipxm}低カロリー\end{CJK}, \begin{CJK}{UTF8}{ipxm}健康オタク\end{CJK} (health freak), \\\begin{CJK}{UTF8}{ipxm}無農薬\end{CJK} (Agrochemical-free), \begin{CJK}{UTF8}{ipxm}ジョギング\end{CJK} (jogging), \begin{CJK}{UTF8}{ipxm}低脂肪\end{CJK} (low fat)\end{tabular} \\ \hline
 Socio-environmental & \begin{tabular}[c]{@{}l@{}}\begin{CJK}{UTF8}{ipxm}温室効果ガス\end{CJK} (Greenhouse gases), \begin{CJK}{UTF8}{ipxm}エコ\end{CJK} (eco), \\\begin{CJK}{UTF8}{ipxm}フェアトレード\end{CJK} (fair trade), \begin{CJK}{UTF8}{ipxm}動物実験\end{CJK} (animal experiment), \\ \begin{CJK}{UTF8}{ipxm}環境保護\end{CJK} (environmental protection), \begin{CJK}{UTF8}{ipxm}偽装食品\end{CJK} (food fraud)\end{tabular}  \\ \hline
 Politics    &  \begin{tabular}[c]{@{}l@{}}\begin{CJK}{UTF8}{ipxm}安倍首相\end{CJK} (Prime Minister Abe), \begin{CJK}{UTF8}{ipxm}ネトウヨ\end{CJK} (online right-wingers), \\\begin{CJK}{UTF8}{ipxm}リベラル\end{CJK} (liberal), \begin{CJK}{UTF8}{ipxm}保守\end{CJK} (conservative), \begin{CJK}{UTF8}{ipxm}ヒラリー\end{CJK} (Hilary), \\\begin{CJK}{UTF8}{ipxm}トランプ\end{CJK} (Trump)\end{tabular} \\ \hline
 \end{tabular}
\end{table}

Figure 2 shows keyword entropy across different categories. 
In Fig. 2A, there is a high degree of interest in major foods, such as ``meat,'' ``fish,'' and ``vegetables.'' 
Comparing the food left-wing and right-wing groups, there is no marked difference in meat and fish, but the right-wing group shows a higher degree of interest in vegetables. 
Of considerable interest here is that the food left-wing showed a greater interest in keywords like ``trans fatty acid'' that can lead to heart disease and ``GM (genetically modified)'' crops such as soybeans and corn in which the produce is manipulated by humans. 
In contrast, the food right-wing showed a higher interest in ``instant noodles,'' the standard-bearer for junk food. 
To confirm in what context these keywords were used, we randomly sampled 200 tweets that included each of these keywords, and then manually checked whether these keywords were used in a positive or negative context. 
It transpired that the food left-wing users mostly made negative statements about trans fatty acid (positive: 1, negative: 199) and genetically modified (GM) crops (positive: 4, negative: 176); instant noodles were mostly used in a positive context by the food-right (positive: 131, negative: 29).
These results align with the image of the food left-wing as liking natural food and the right-wing as liking fast food.

\begin{figure}[t]
\centering
\includegraphics[clip,width=1.0\textwidth]{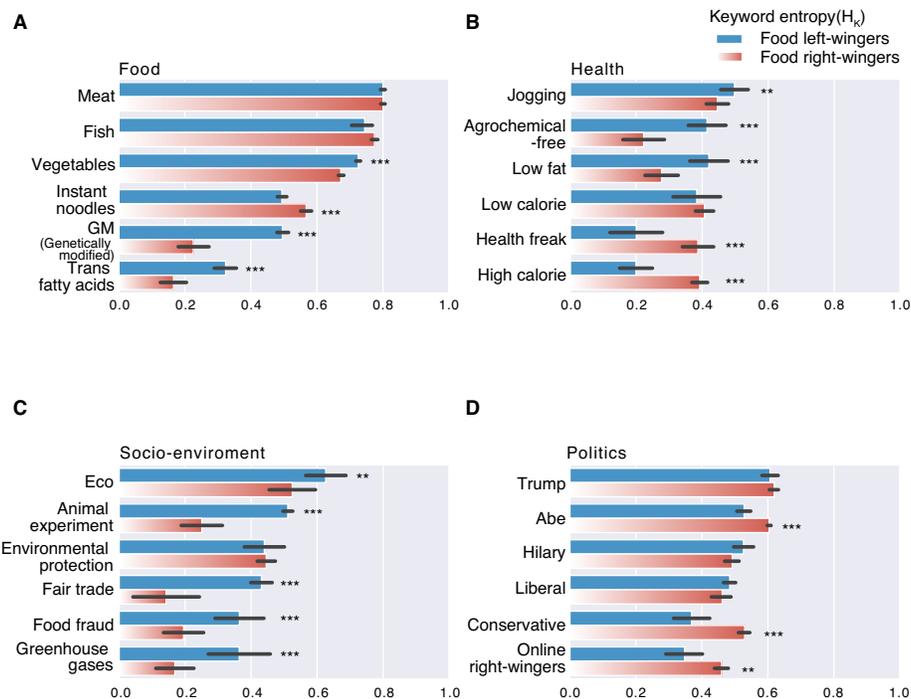}
 \caption{Degree of collective interest in keywords related to food, health, socio-environment, and politics (means and SDs computed from the bootstrapped samples).}
 \label{fig2}
\end{figure}

Next, we examined the two groups to determine their preferences for categories other than food. 
This can be confirmed in Figs. 2B-D when looking at categories of health, socio-environment, and politics. 
In the health category, the food left-wing had a strong interest in ``jogging,'' ``agrochemical-free,'' and ``low fat'' while the food right-wing had a strong interest in ``high calorie,'' and ``health freak.'' 
This brings to mind the image of food right-wing users as eating fast food while worrying about the calories associated with it. 
Given that food left-wing users show a greater interest in most of the keywords in the socio-environmental category, we can see how this concurs with the image of food left-wing users as highly conscious about socio-environmental issues. 
Interestingly, the food right-wing showed a strong interest in keywords like ``conservative'' and ``online right-wingers'' and ``Abe'' (the current prime minister of Japan, seen as a right-wing politician). 
Thus, there is some degree of correlation between food and politics.
In contrast, such correlation was not observed in the food left-wing. 

Figure 3 summarizes the results of the keyword entropy when plotted by the laterality index ($LI$), showing specific differences between the two groups in categories that go beyond food. 
Again, a greater $LI$ implies a greater degree of interest among food right-wing users, while a smaller $LI$ implies a greater degree of interest among food left-wing users. 
We will not repeat the same observations shown in Fig. 2. 
Rather, we emphasize that the food left-wing group had a marked interest in socio-environmental issues and the food-right showed some interest in political issues. 

\begin{figure}[t]
\centering
 \includegraphics[clip,width=1.0\textwidth]{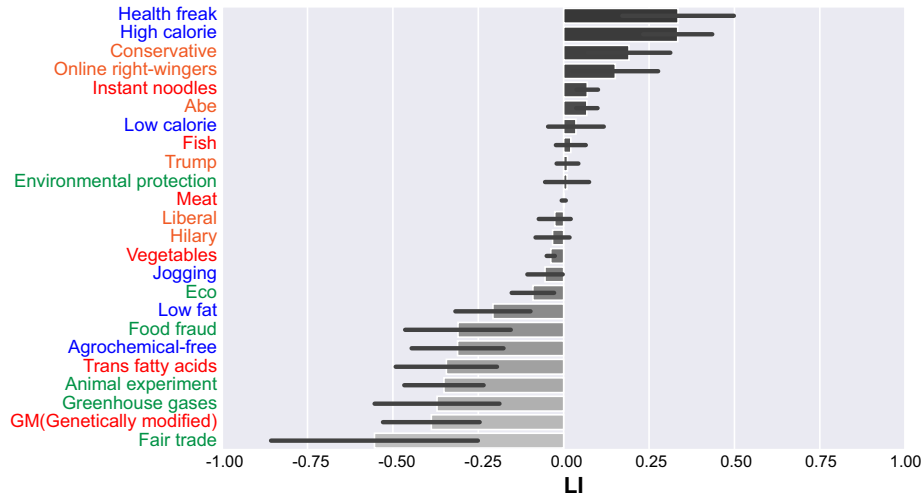}
 \caption{Collective interest in all keywords (means and SDs computed from the bootstrapped samples). In the PDF version of this paper, colors correspond to categories (Food: red, Health: blue, Socio-environment: green, Politics: orange). }
 \label{fig3}
\end{figure}

It is notable that the term ``animal experiment'' is of high interest among the food left-wing. 
Using association networks~\cite{Sasahara2016}, we examined whether this is indeed the case. 
Figure 4 is an association network of the words obtained when using ``animal experiment'' as the seed word when mapping against food left-wing users. 
The links show word associations with a cosine similarity of 0.4 or higher. 
The term ``animal experiment'' directly ties to ``animal cruelty'' and then to word clusters comprising terms like ``cruel,'' ``torture,'' ``fur,'' and ``mink.'' 
Other notable connections seeded from ``animal experiment'' include ``boycott,'' ``cosmetics,'' and ``military research.'' 
This association network reveals how food left-wing users harbor a strong negative image toward animal experiments. 
We conducted the same process on the food right-wing data, but there was no markedly negative association with the term ``animal experiment.''

\begin{figure}[t]
\centering
 \includegraphics[clip,width=0.9\textwidth]{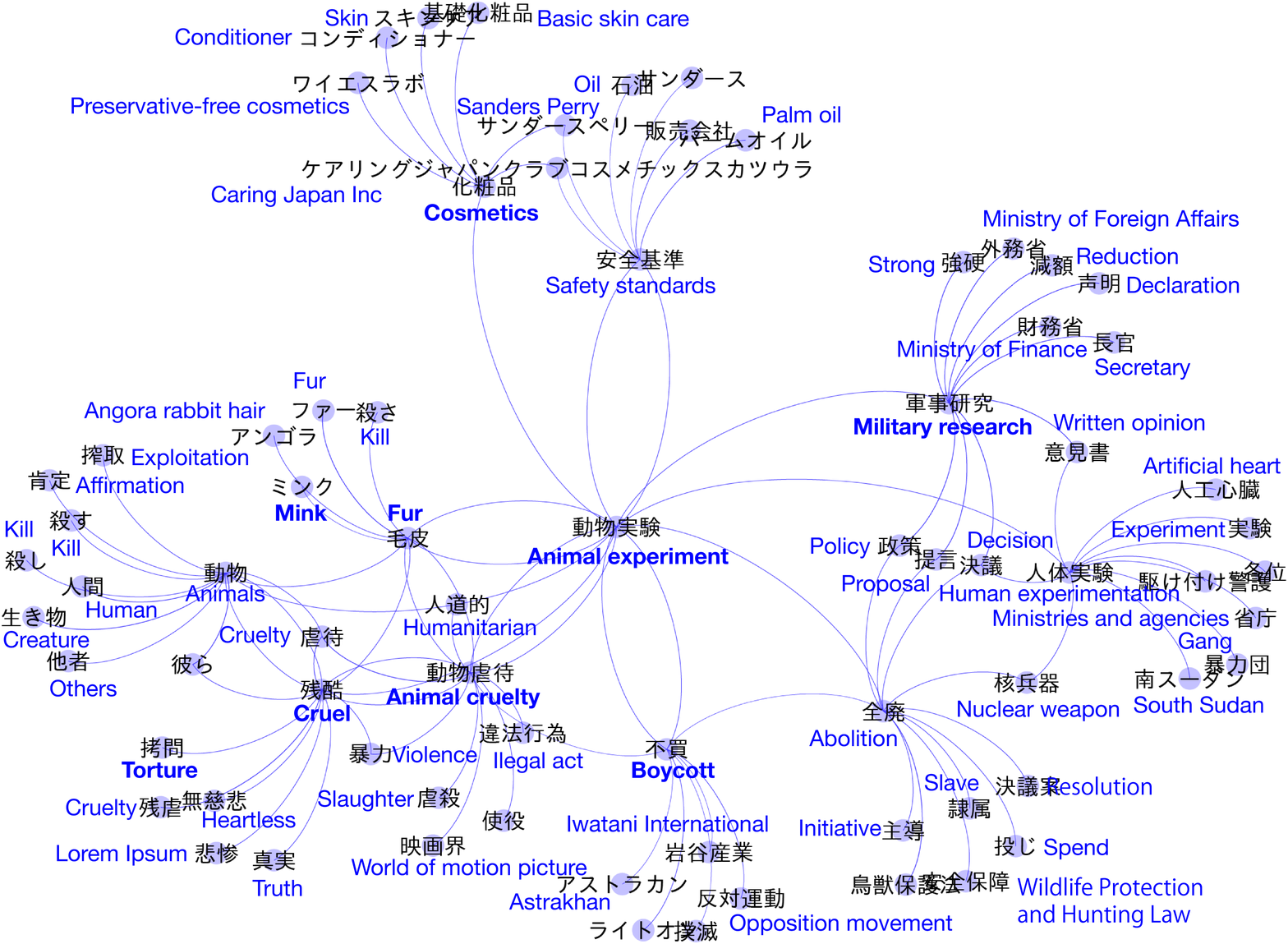}
 \caption{Association network of the term “animal experiment” for food left-wing users.}
 \label{fig4}
\end{figure}

These findings indicate that the food left-wing has strong feelings about animal welfare and strongly shirks from the idea of artificially manipulating animals. Therefore, we can plainly see the way in which consumer attitudes, such as refusing to buy cosmetics that have been tested on animals or refusing to buy mink coats that are the result of killing and skinning animals, are manifested in tweets. 

\subsection*{Consumer interests in brands}

Next, we present an investigation into consumer interests in specific brands. 
We used a 2016 Forbes survey \cite{Forbes2016} and a 2016 Nikkei survey \cite{Nikkei2016} to identify 24 global brands across four industries (Table 3). 
The keyword entropy ($H_{k}$) is measured based on the co-occurrence of brand names and positive words (Table 4) in Dataset 2. 
In this case, a larger $H_{k}$ implies more positive interest in the brand.

\begin{table}[h]
\caption{Industries and major brands used}
\begin{tabular}{|l|l|}
\hline
Technology  & Apple, Google, Microsoft, Facebook, Amazon, Sony        \\ \hline
Retail      & McDonalds, Starbucks, IKEA, Costco, KFC, UNIQLO         \\ \hline
Beverage   & Pepsi, Coca-Cola, Corona, Red Bull, Budweiser, Heineken \\ \hline
Automobile & Toyota, BMW, Mercedes, Honda, Volkswagen, Nissan        \\ \hline
\end{tabular}
\end{table}

\begin{table}[h]
\caption{Positive keywords}
 \begin{tabular}{|l|}
 \hline
 \begin{tabular}[c]{@{}l@{}}\begin{CJK}{UTF8}{ipxm}ほしい/欲しい\end{CJK} (want), \begin{CJK}{UTF8}{ipxm}買った/購入した\end{CJK} (bought), \begin{CJK}{UTF8}{ipxm}格好いい\end{CJK} (cool), \begin{CJK}{UTF8}{ipxm}良い\end{CJK} (nice),\\ \begin{CJK}{UTF8}{ipxm}すばらしい/素晴らしい\end{CJK} (awesome), \begin{CJK}{UTF8}{ipxm}好き\end{CJK} (like), \begin{CJK}{UTF8}{ipxm}美味い/うまい\end{CJK} (tasty),\\ \begin{CJK}{UTF8}{ipxm}美味しい/おいしい\end{CJK} (delicious), \begin{CJK}{UTF8}{ipxm}ほしい/欲しい\end{CJK} (want),
 \begin{CJK}{UTF8}{ipxm}素晴らしい/すばらしい\end{CJK} (excellent), \\\begin{CJK}{UTF8}{ipxm}素敵\end{CJK} (sweet), \begin{CJK}{UTF8}{ipxm}最高\end{CJK} (best)\end{tabular} \\ \hline
 \end{tabular}
\end{table}

Figure 5 shows the degree of positive interest in brands across the categories of technology, retail, beverage, and automobile, comparing between the food left-wing and food right-wing.
Figure 6 summarizes these results in terms of laterality index (\emph{LI}). 

\begin{figure}[t]
\centering
 \includegraphics[clip,width=1.0\textwidth]{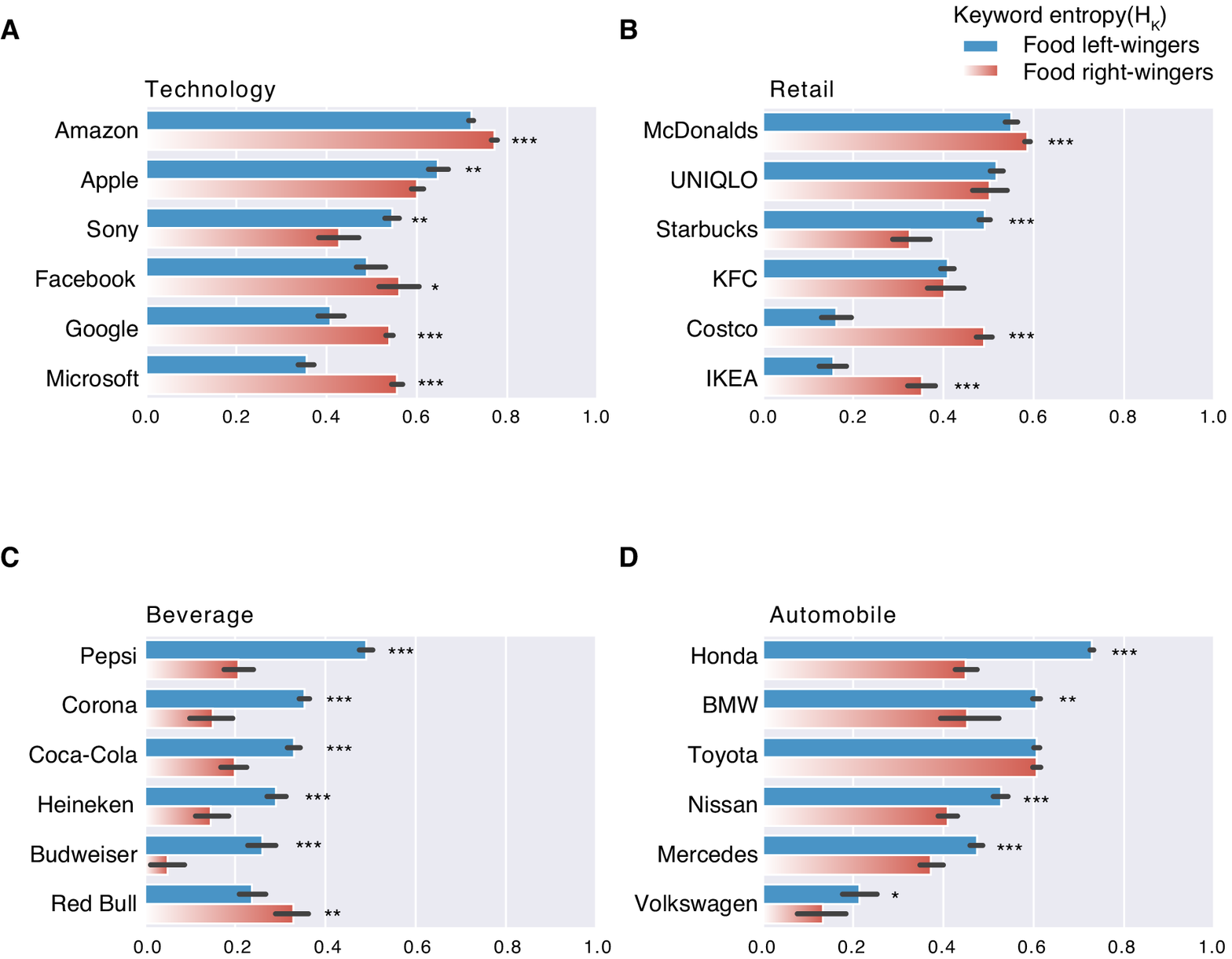}
 \caption{Positive interest in brands related to technology, retail, beverage, and automobile (means and SDs computed from the bootstrapped samples).}
 \label{fig5}
\end{figure}

\begin{figure}[h]
\centering
\includegraphics[clip,width=1.0\textwidth]{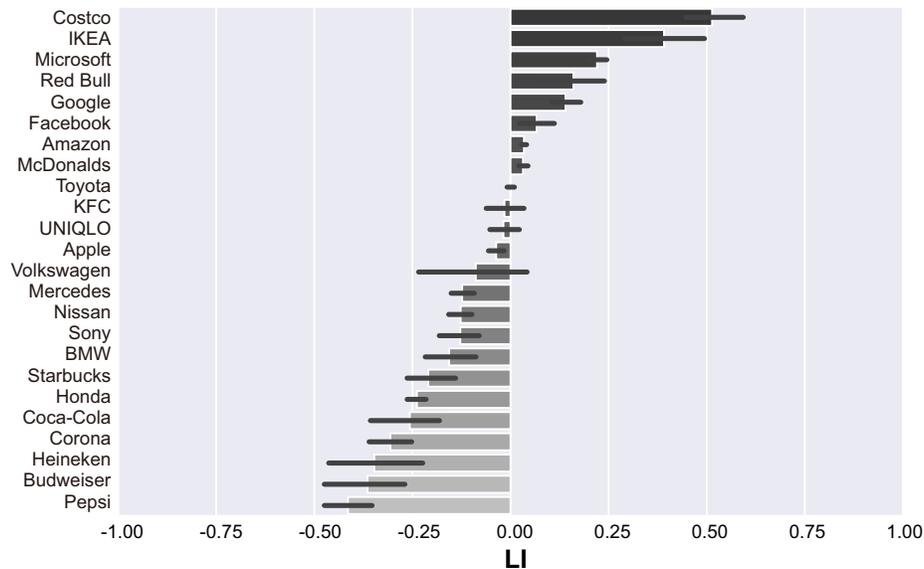}
 \caption{Positive interest in brands (means and SDs computed from the bootstrapped samples).}
 \label{fig6}
\end{figure}

The food right-wing group preferred most of the technology firms except for Apple and Sony~\footnote{Given that most posts about Amazon concern online shopping, Amazon might be more appropriately interpreted belonging to the retail category rather than technology.}. 
In the retail space, the food right-wing showed a markedly higher interest in IKEA and Costco. 
Incidentally, Starbucks---which we mentioned in the introduction---showed an expected high level of interest among the food left-wing. 
Further differences were seen in the beverage space where the food left-wing showed strong interest in overseas beer brands like Corona, while the food right-wing showed high interest in energy drink brands like Red Bull. 
A counter intuitive finding is that the food left-wing's positive interest in soda, especially in Pepsi.  
We observed positive tweets, such as ``Pepsi is way tastier than Coke. As for Guarana, it's a bit in the middle...'' 
The food-left wing showed a higher interest prefers in most of the automobile brands except Toyota.

The differences in positive interest between the food left-wing and right-wing groups are a useful metric when determining which products to promote through advertising based on the unique preferences of each group.

\subsection*{Network of social interactions in food left-wing and right-wing users}
As described previously, we created a retweet network ($G$) based on the propagation of retweets ($|V|$=165,609, $|E|$=382,899). 
Figure 7 visualizes the largest connected components with a maximum of 50 or more orders of connection
($|V|$=1,113, $|E|$=33,231). 
Nodes represent users with blue being the food left-wing and red being the food right-wing. 
Yellow represents those whose orientation could not be ascertained (i.e., users other than the left-wing and right-wing users in Dataset 2). 
The links represent retweet transmissions. 
One salient feature of these networks is the formation of two distinct left and right clusters. 
Unlike the food left-wing users scattered on the right of Fig. 7, those on the left side of the graph suggest that the food-left users actively communicate about organic food-related information via retweets, thereby forming a cohesive cluster.
In contrast, the food right-wing does not manifest as densely as the food left-wing, suggesting retweet communications are less cohesive.

\begin{figure}[t]
\centering
 \includegraphics[clip,width=1.0\textwidth]{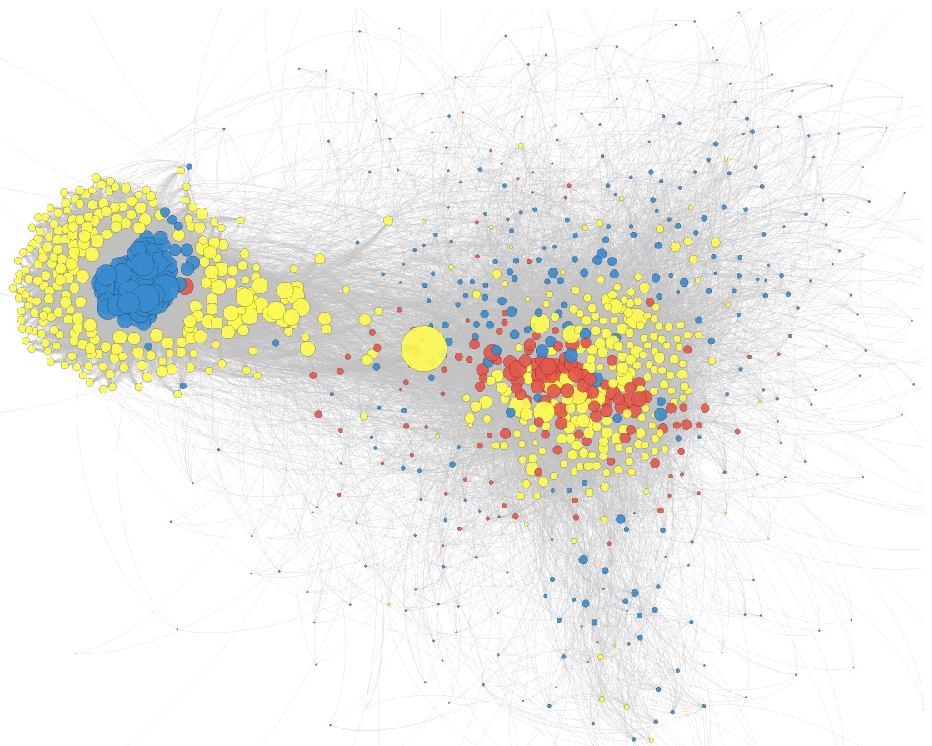}
 \caption{Retweet networks of food left-wing and right-wing users. Nodes denote users. Blue corresponds to the food left-wing, and red corresponds to the food right-wing; yellow is unknown. The node size is proportional to overall degree. The links denote retweet transmissions.}
 \label{fig7}
\end{figure}

Another notable feature in Fig. 7 is the existence of a ``bridge'' (the large yellow node) that links two separate clusters.
The bridge has major potential influence because this class of users can convery information to both the food left-wing and right-wing users. 
This account belongs to livedoor News (@livedoornews) that manually curate and reposte articles posted to the livedoor News website. 
Since livedoor News has a food section, this account makes many posts about food and cuisine; some are preferred by the food left-wing and others by the food right-wing. 

\section*{Discussion}
We have demonstrated differences in personal values and consumer awareness for left-wing and right-wing users via social data analysis. 
Food preferences are not a binary of left or right but rather a continuum, as personal attributes are multi-dimensional in nature. 
This research, however, implies that mapping the plurality of personal attributes across the single dimension of left and right and looking at these two extremes does offer useful insights.
Measuring the degree of collective interest in certain keywords and brand names revealed that beyond the domain of food, the food left-wing has a strong interest in socio-environmental issues (notably, animal experiments) and the food right-wing has a higher interest in large-scale shopping malls offering discounts and volume sales of goods, such as IKEA and Costco. 
In addition, we observed differences between the food left-wing and right-wing groups in word usage.
Table 5 shows the top 100 popular keywords ranked by the average TF-IDF score~\cite{manning2008introduction}. 
Note that the word's TF-IDF score is its term frequency divided by its document (tweet) frequency. 
Many food left-wing related words are ranked in the table in the food left-wing group (e.g., beauty, health, and nature), while several food-right wing related words are ranked in the table in the food right-wing group (e.g., supermarket and ice cream).

\begin{table}[h]
\caption{Top 100 popular keywords ranked by the average TF-IFD socre.}
\centering
\includegraphics[clip,width=1.0\textwidth]{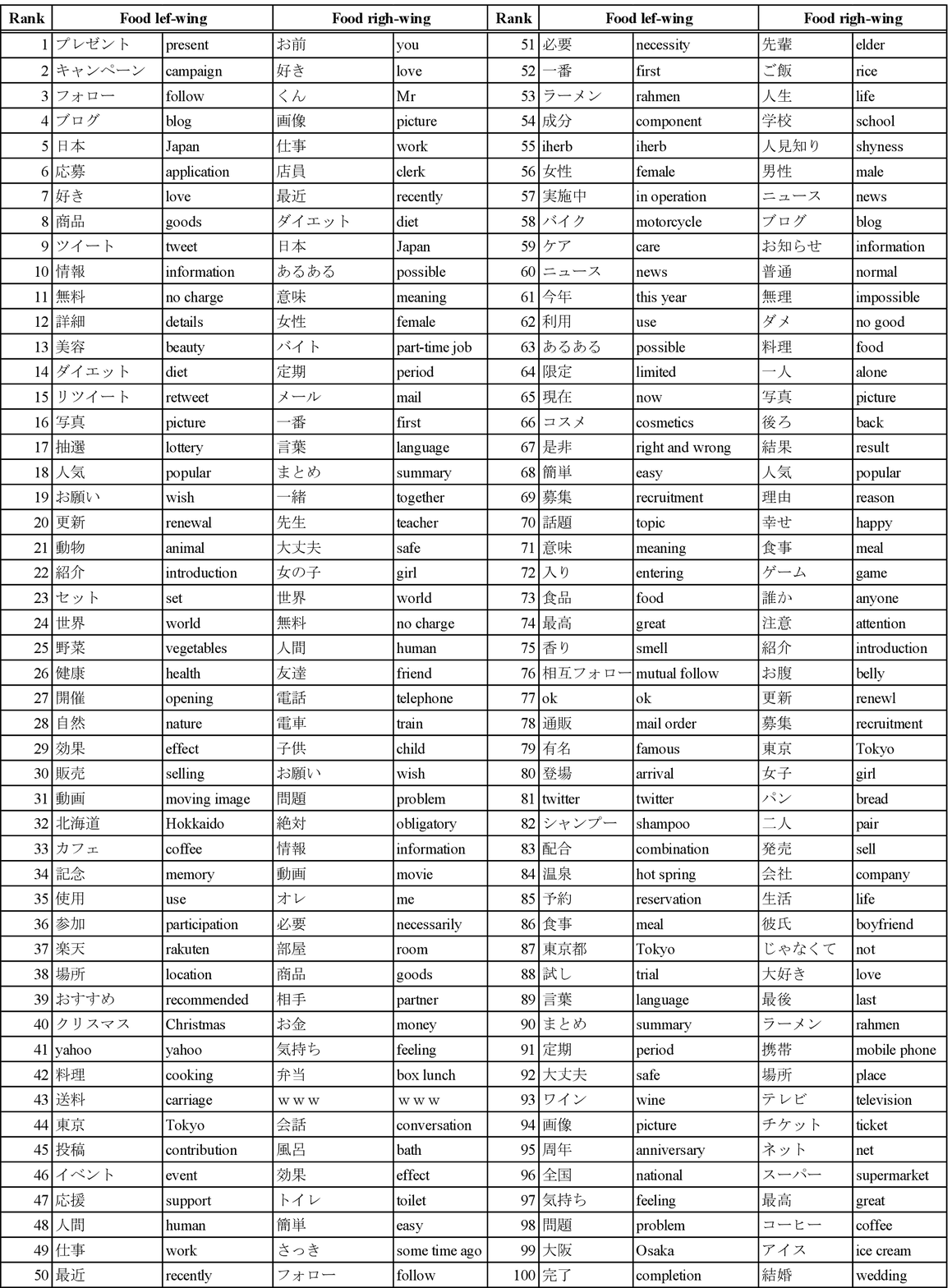}
\end{table}

The segregated network structures in retweet interactions revealed that food-related information is also consumed differently in the food left-wing and right-wing groups. 
One point of interest in these networks is that clusters of food left-wing and right-wing users are linked via the conduit of a news outlets (livedoor News). 
Given that online social networks are mechanisms for disseminating information, identifying users' food preferences might be efficient in the distribution of a specific piece of information, an advertisement, or a campaign.

Looking solely at these results, one could conjecture as to the differences in consumer attitudes among these groups---the food-left wing rejects industrialization and prefers a return to nature, and the food right-wing does the opposite. 
However, reality is not quite so simple. 
Indeed, the food left-wing had a higher degree of interest in Apple and Sony as technology brands, as well as in most automobile brands. 
Therefore, we cannot say conclusively that the food left-wing comprehensively rejects industrialization and prefers a return to nature in contexts other than food. 

This study has implications for social sciences and applications.
The quantification of personal attributes is an important procedure in many areas of social sciences. 
Personal attributes, however, are difficult to accurately measure by survey research alone. 
Given the fact that food identity could be a concise yet useful proxy for various personal attributes and that there are numerous reports about food on social media, mining food identities from online digital traces can contribute to social science research.
Furthermore, the idea of food identity could be applicable in social media marketing and other applications.
For example, rather than relying on influencers (\cite{Keller2003}), which represent a small and limited population, or on simply trafficking large quantities of advertisements at random, selectively running advertisements based on the underlying values around food is much more likely to hit the mark.
However, one should remember that simply extrapolating the concepts of the food left-wing and right-wing without conducting prior research would prove detrimental when doing social sciences or engaging in marketing.

We recognize the limitations of social media analysis for studying food identity, because social media users are biased toward age, gender, geolocation, etc. 
To supplement social data analysis, we have to incorporate survey research, which is one of our future research directions.
Moreover, the keyword-based food preference identification done here has potential concerns with misclassification. 
For example, food left-wing users who often post criticisms about fast food could be classified into food right-wing users; similar errors could occur for food right-wing users. 
In addition, this approach cannot capture criticisms and cynicisms, potentially conflating very different types of contexts.
To resolve these issues, contextual information needs to be considered to improve the classification accuracy.
Although several limitations remain, our social media analysis has demonstrated that food identity can be a useful concept to address personal attributes and consumer awareness. 

\section*{Acknowledgements}
K.S. thanks to M. Karasawa and K. Hioki for discussions. This research was supported in part by Yoshida Hideo Memorial Foundation, JSPS/MEXT KAKENHI Grant Numbers JP16K16112 and JP17H06383 in \#4903, JST PRESTO Grant Number JPMJPR16D6, and JST CREST Grant Number JPMJCR17A4.


 \end{document}